\documentclass[12pt]{article}
\usepackage{amssymb}
\usepackage{amsfonts}
\usepackage{amsmath}
\usepackage{latexsym}
\usepackage[T1]{fontenc}
\usepackage[utf8]{inputenc}
\usepackage{lmodern}
\usepackage{latexsym}
\usepackage{graphicx}

\begin{document}
\title{\bf{Uncertainty relations: curiosities and inconsistencies }}
\author{
K. Urbanowski\footnote{e--mail:  K.Urbanowski@if.uz.zgora.pl, $\;$ k.a.urbanowski@gmail.com}\\
University of Zielona G\'{o}ra, Institute of Physics, \\
ul. Prof. Z. Szafrana 4a, 65--516 Zielona G\'{o}ra, Poland.
}


\maketitle

\begin{abstract}
Analyzing general uncertainty relations one can find
that there can exist such pairs
of non-commuting observables $A$ and $B$ and such vectors that the lower bound for
the product of standard deviations  $\Delta A$ and  $\Delta B$ calculated for these vectors is zero: 
$\Delta A\,\cdot\,\Delta B \geq 0$. Here we discuss examples of such cases and some other inconsistencies
which can be found performing a rigorous analysis of the  uncertainty relations in  some special cases.
As an illustration of such cases matrices $(2\times 2)$ and $(3 \times 3)$  and the position--momentum uncertainty relation for a quantum  particle in the box are considered.
The status of the uncertainty relation in $\cal PT$--symmetric quantum theory and the problems associated with it are also studied.

\end{abstract}

\section{Introduction}
The famous Heisenberg uncertainty relations \cite{H,H2} play an important and significant  role in the understanding of the quantum world and in explanations of its properties.
There is a mathematically rigorous derivation of the position--momentum uncertainty relation
and the uncertainty relation for any pair of non--commuting observables, say $A$ and $B$, within the standard formalism of Schr\"{o}dinger and von Neumann \cite{Robertson,Schrod-1930,M}.
Among physicists who do not deal with theoretical research on the mathematical foundations of quantum mechanics, and in particular with a rigorous  derivation of the uncertainty principles, there is an almost common belief
based on the textbooks treatment of the problem (see eg.  \cite{Mr,Gr}) that
if one has a pair of non-commuting observables $A$ and $B$ then the  the product of standard deviations  $\Delta A$ and  $\Delta B$ calculated for them is always large than some nonzero positive number, say $c$:
\begin{equation}
\Delta A \,\cdot\, \Delta B \geq c> 0. \label{ur-c}
\end{equation}
Here we show that such a belief may lead to confusions:
It appears
that there may exist such vectors that the lower bound for this product is zero. Simply,
there exist such pairs of non--commuting operators $A$ and $B$ and such vectors from the Hilbert state space that
for the standard deviations calculated for these vectors there is
$\Delta A\,\cdot\,\Delta B \geq 0$ (see, e.g. \cite{mpla-2020}).
The motivation of the paper is
to examine such and similar cases and to discuss other
limitations of Robertson--Schr\"{o}dinger uncertainty relation (\ref{ur-c}) and
inconsistencies  as well as mathematical problems connected with
this relation.
Here we show examples of the cases where one can find that  there is $\Delta A\,\cdot\,\Delta B \geq 0$ for some vectors although $[A,B] \neq 0$.
The simplest cases are illustrated using Pauli matrices nad Gell--Mann matrices. One meets a  much more complicated situation in the case of a problem of a quantum particle in the box with  perfectly reflecting and impenetrable walls: In this case we analyze the position--momentum uncertainty relation. We show that this problem leads to some paradoxical situations and  generates some inconsistencies.
 The solution of these inconsistencies is proposed: From the point of view of the classical mechanics  the particle in the box is a constrained system and  the use of the  position operator consistent with the constraints can solve these inconsistencies. Analyzing the problem of particle in the box we observed that
 some subtle properties of such system depending on the choice of the boundary conditions
 may be related to the symmetry properties of the problem under study. For this reason, we have attempted to investigate the problem of the uncertainty relations in $\cal PT$--symmetric quantum mechanics. We found that within $\cal PT$ symmetric quantum mechanics a relation corresponding to the uncertainty relations discussed, e.g. in \cite{H,H2,Robertson,Schrod-1930} may not exist for every pair of non--commuting operators. We also found that if it exists for a pair of noncommuting $\cal PT$--symmetric observables than it can not be considered as universally valid.

  The paper is organized as follows: In Section 2  the reader finds  some preliminaries. The case of Pauli and Gell--Mann matrices  is analyzed in Sec. 3. Section 4 contains analysis of the case of
  a quantum particle in the box with  perfectly reflecting and impenetrable walls. Discussion of the problem of uncertainty relations in $\cal PT$--symmetric quantum theory is presented in Sec. 5.
   Sec. 6 contains a discussion and conclusions.

\section{Preliminaries}

The uncertainty principle was formulated by Heisenberg \cite{H,H2} for the position and momentum and it
can be written as follows
\begin{equation}
\Delta_{\phi} X\ \cdot \Delta_{\phi} P_{x}\,\geq \,\frac{\hbar}{2}. \label{H1}
\end{equation}
Heisenberg considered $\Delta_{\phi} X$ and $\Delta_{\phi} P_{x}$ as {\em "precisions"} with which the values $x$ and  $p$ are known \cite{H}.
Practically from the moment of the publication of Heisenberg's works \cite{H,H2}, the ongoing discussion on how to interpret the inequality (\ref{H1}) began (see, eg. \cite{Werner,Busch,Busch1}).

The contemporary interpretation of  $\Delta_{\phi} X$ and $\Delta_{\phi} P_{x}$
considered in this paper
comes from the derivation of the uncertainty relation made by Robertson
\cite{Robertson} and Schr\"{o}dinger \cite{Schrod-1930},  (see also \cite{M}):
$\Delta_{\phi} X$ and $\Delta_{\phi} P_{x}$ denote the  standard (root--mean--square) deviations or variances. In a general case for an self--adjoint operator  $F$ acting in ${\cal H}$ the standard deviation is defined as follows
\begin{equation}
\Delta_{\phi} F = \| \delta F|\phi\rangle\|, \label{dF}
\end{equation}
where $\delta F = (F - \langle F\rangle_{\phi}\,\mathbb{I} )$, and $\langle F\rangle_{\phi} \stackrel{\rm def}{=} \langle \phi|F|\phi\rangle$ is the average (or expected) value of $F$ calculated for the normalized vector $|\phi\rangle \in {\cal H}$, provided that $|\langle\phi|F|\phi \rangle |< \infty$.
(Note that from the definition of $\delta F$ it follows that $\delta F$ must be the self--adjoint operator if $F$ is self--adjoint).
The equivalent definition is:  $\Delta_{\phi} F \equiv \sqrt{\langle F^{2}\rangle_{\phi} - \langle F\rangle_{\phi}^{2}}$.
(In Eq. (\ref{H1})  $F$ denotes position  and momentum operators $x$ and $p_{x}$ as well as  their squares).
Within the quantum theory the operator $F$ represents observable $F$.
So, the uncertainty
principle is a relation connecting standard deviations (variances) calculated for a pair of non--commuting observables (that is, self--adjoint operators) acting in a Hilbert space ${\cal H}$.
 In general, relations (\ref{ur-c}) and  (\ref{H1}) results from basic assumptions of the quantum theory and from the geometry of Hilbert space \cite{Teschl}.
 Relations having the form (\ref{ur-c})
 hold for any two observables, say $A$ and $B$, represented by non--commuting self--adjoint operators $A$ and $B$ acting in the Hilbert space of states (see \cite{Robertson} and also \cite{Schrod-1930}), such that $[A,B]$ exists and $|\phi\rangle \in {\cal D}(AB) \bigcap {\cal D}(BA)$, (${\cal D}({\cal O})$ denotes the domain of an operator $\cal O$ or of a product of operators):
\begin{equation}
\Delta_{\phi} A \cdot \Delta_{\phi} B\;\geq\;\frac{1}{2} \left|\langle [A,B] \rangle_{\phi} \right|.\label{R1}
\end{equation}

 As it was said in the general case the relation (\ref{R1}) results from the geometry of the Hilbert space, strictly speaking from the Schwartz inequality: Let
$|\psi_{1}\rangle, |\psi_{2}\rangle \in {\cal H}$, then
one has
$| \langle \psi_{1}|\psi_{2}\rangle |\, \leq \, \left\| \, |\psi_{1}\rangle \right\| \; \left\|\, |\psi_{2}\rangle \right\|$.
Next taking
$|\psi_{1} \rangle =  \delta A |\phi\rangle$ and $ |\psi_{2}\rangle =  \delta B |\phi\rangle$,
after some algebra one obtains the inequality (\ref{R1}) --- details can be found in Section 2 in \cite{mpla-2020} and in many textbooks  and journal articles.
Now if to identify operators $A$ and $B$ acting in the Hilbert space ${\cal H} = L^{2}(\mathbb{R})$: $A$ with the momentum operator, $P_{x}$, in quantum theory,  $B$ with the position operator $X$, and then using the commutation relation,
\begin{equation}
[P_{x},X]= - i \hbar\,\mathbb{I}, \label{[PX]}
\end{equation}
one obtains from (\ref{R1}) the inequality (\ref{H1}), i.e.  the Heisenberg uncertainty relation.

Note that starting with the Schwartz inequality
all subsequent calculations and transformations leading to the result (\ref{R1})  are purely mathematical operations and there is no physics in them (see, e.g. \cite{mpla-2020}):
The inequality  (\ref{R1}) is a purely mathematical inequality and examining when and for which vectors it occurs and for which it does not occur is a mathematical task.  Physics will appear only when physical quantities are assigned to operators $ A$ and $B$ and the Hilbert space on which they act is identified with the space of the states of the  physical system considered.

As it was mentioned,
there is still a discussion on how to interpret  inequalities (\ref{R1}) and (\ref{H1}) and how to improve them (see, e.g. \cite{Werner} and references therein, \cite{Busch,Busch1,Fol,Cow,Dou,Park,Benitez,Ozawa,Dias} and many other papers).
From the derivation of the formula (\ref{R1}) it follows that the standard deviations $\Delta_{\phi}A$  and $\Delta_{\phi}B$ characterize the statistical distribution of the most probable values of $A$ and $B$ in the state $| \phi\rangle$. The inequality (\ref{R1}) does not depend on a possible influence of the measuring device on the result of measurements and on the statistical distribution of values of $A$ and $B$  measured by this device. So, it seems that
a  safe interpretation of (\ref{R1}) is the interpretation  close to that one can find in \cite{Ken}, namely that it is impossible to prepare a system in a state $|\phi\rangle$ that non--commuting observables $A$ and $B$ have both  their probability distributions of values of $A$ and $B$ in this state sharply concentrated around a single value (see, \cite{Werner,Busch,Busch1}).
Therefore the relation (\ref{R1}) is sometimes called the {\em preparation uncertainty relation} \cite{Werner,Busch,Busch1,Benitez,Ozawa}.
There is also another, probably the most popular interpretation of inequality (\ref{H1}) in the literature. Namely,
Heisenberg’s relation (\ref{H1}) is considered  as a trade-off between the precision
$\Delta_{\phi}X$ of an approximate position measurement and the momentum disturbance
$\Delta_{\phi}P$ incurred by that measurement (see, eg. \cite{Werner,Ozawa}). This is the {\em error--disturbance} or {\em noise--disturbance uncertainty relation} (see, eg. \cite{Ozawa,Ozawa1,Renes,Busch2,Busch3,Ciril}. The relation (\ref{R1}) can be understood analogously. The proof of this relation having similar form to the relations (\ref{H1}), (\ref{R1}) can be found, e.g. in \cite{Ozawa,Ozawa1,Busch2}.
One more interpretation of the uncertainty relation can be found in the literature. It is so--called
{\em Heisenberg uncertainty relation for joint measurements}. It can be
 generally formulated
as follows \cite{Ozawa1}: For any apparatus $\mathbb{A}$ with two outputs for
the joint measurement of $ A$ and $B$, the relation (\ref{R1})
holds for any input state $|\phi\rangle$, where, in this case  $\Delta_{\phi}A$ is replaced by $\epsilon(A,\phi,  \mathbb{A})$, $\Delta_{\phi}B \rightarrow \epsilon(B,\phi,  \mathbb{A})$ and $\epsilon(X,\phi,  \mathbb{A})$
stands for the noise of the $X$ measurement in state $|\phi\rangle$ using apparatus $\mathbb{A}$ for
$X= A,B$ \cite{Ozawa1,Ciril,Werner1,Ozawa2}. The proof of this relation can be found, e.g. in  \cite{Ozawa2}. It requires the assumption that the experimental mean values of the outcome $\bf x$ of the $A$ mesurement and the outcome $\bf{y}$ of the $B$ measurement should coincide with the mathematical expectation values  of observables $A$ and $B$, respectively, on any input state $|\phi\rangle$ \cite{Ozawa2}. So, due to such an assumption the final form of the {\em uncertainty relation for joint measurements} is
analogous to that given by the inequality (\ref{R1}). In  general,
 a common feature of all these cases is that the uncertainty relation takes the form  considered in this paper, that is the form given by inequalities (\ref{ur-c}), (\ref{H1}), (\ref{R1}).
A discussion of different aspects of these interpretations as well as attempts to improve uncertainty relations are still continued and can be found  in many papers (see, e.g.
\cite{Furrer,Dam,Coles,Thek,Werner2}).

In this paper attention will be focused on the definition of standard deviations given by the formula (\ref{dF}) and properties of (\ref{R1}) resulting from this definition.
It has been pointed out in \cite{mpla-2020} that it is not necessary for $A$ and $B$ to commute, $[A,B] =0$, in order that  $\langle\phi|[A,B]|\phi \rangle = 0$ for some $|\phi\rangle \in {\cal H}$. Simply it may happen that for some $|\phi \rangle \in {\cal H}$ and for some non-commuting operators $A$ and $B$ the expectation value of the commutator $[A,B]$ vanishes:  $\langle\phi|[A,B]|\phi \rangle = 0$ and then  the inequality (\ref{R1}) takes the following form:
\begin{equation}
\Delta_{\phi} A \cdot \Delta_{\phi} B\;\geq\;0. \label{dAdB=0}
\end{equation}
This means that in such cases the inequality (\ref{R1}) having the form (\ref{dAdB=0})
does not impose  any restrictions for the values of $\Delta_{\phi} A$ and $ \Delta_{\phi} B$ besides the condition that there should be  $0 \leq \Delta_{\phi}A < \infty$ and  $0 \leq \Delta_{\phi}B < \infty$. Examples of such and similar cases will be analyzed in the next Section.

\section{Simple algebraic examples}

Here we present examples of self--adjoint operators (matrices) for which the inequality (\ref{R1}) has the form (\ref{dAdB=0}). So, let us considerer for a start the simplest case of $(2\times2)$ matrices. Using Pauli matrices
\begin{equation}
\sigma_{x} = \left(
               \begin{array}{cc}
                 0 & 1 \\
                 1 & 0 \\
               \end{array}
             \right),\;\;
\sigma_{y} = \left(
               \begin{array}{cc}
                 0 & -i \\
                 i & 0 \\
               \end{array}
             \right),\;\;
\sigma_{z} = \left(
               \begin{array}{cc}
                 1 & 0 \\
                 0 & -1 \\
               \end{array}
             \right), \label{pauli}
                         \end{equation}
one has $\sigma_{x} = \sigma_{x}^{+},\;\sigma_{y} = \sigma_{y}^{+},\;\sigma_{z} = \sigma_{z}^{+}$ and
\begin{equation}
[\sigma_{x},\sigma_{y}] = 2i \sigma_{z}.
\end{equation}
Identifying $\sigma_{x},\;\sigma_{y}$ with operators $A$ and $B$ respectively one can rewrite (\ref{R1}) as follows
\begin{equation}
\Delta_{\phi} \sigma_{x}\,\cdot\,\Delta_{\phi} \sigma_{y} \geq \frac{1}{2}\left|\langle [\sigma_{x},\sigma_{y}] \rangle_{\phi} \right| \equiv
\left|\langle \sigma_{z} \rangle_{\phi}\right|, \label{R1-sigma}
\end{equation}
where
\begin{equation}
|\phi\rangle = N \left(
                 \begin{array}{c}
                   a \\
                   b \\
                 \end{array}
               \right),
\end{equation}
$ N = \left(|a|^{2} + |b|^{2}\right)^{-1/2}$,
$\;a,b \in \mathbb{C}$,  $\langle \sigma_{x} \rangle_{\phi} = \langle \phi |\sigma_{x}|\phi \rangle = 2 N^{2}\,\Re\,[a^{\ast}b]$ and
 $(\Delta_{\phi} \sigma_{x})^{2} = \langle \phi |\sigma_{x}^{2}|\phi\rangle - \langle\sigma_{x}\rangle_{\phi}^{2} \equiv 1 - 4\,N^{4} \,(\Re\,[a^{\ast}b])^{2}$, and so on.
(Here $\Re[z]$ and $\Im[z]$ denote real and imaginary parts of $z$ respectively).
It is easy to see that $\langle \sigma_{z} \rangle_{\phi} = \langle \phi |\sigma_{z}|\phi \rangle = N^{2}\,\left( |a|^{2} - |b|^{2} \right)$ which means that
$\left|\langle \sigma_{z} \rangle_{\phi}\right| > 0$
if $|a| \neq |b|$. Choosing
$|\phi\rangle$ such that    $|a| = |b|$, e.g., $a=b =1$,
\begin{equation}
|\phi\rangle \;\Rightarrow\; |\phi_{1}\rangle = \frac{1}{\sqrt{2}}\, \left(
                 \begin{array}{c}
                   1 \\
                   1 \\
                 \end{array}
               \right),
\end{equation}
 one finds that $\left|\langle \sigma_{z} \rangle_{\phi_{1}}\right| \equiv 0$,
 and, as a result the
 inequality  (\ref{R1-sigma}) will take the form of  (\ref{dAdB=0}) for $|\phi\rangle_{1}$.
  We have $\langle \sigma_{y} \rangle_{\phi}  = 2 N^{2}\,\Im\,[a^{\ast}b]$ and
 $\Delta_{\phi}\sigma_{y} = 1 -4 N^{4} (\Im\,[a^{\ast}b])^{2}$. This means that for  $|\phi_{1}\rangle $ one obtains $\langle \sigma_{y} \rangle_{\phi_{1}} = 0$ and   $\Delta_{\phi_{1}}\sigma_{y} = 1$.
 Note that in this case  $\Delta_{\phi_{1}} \sigma_{x} = 0$ because the vector $|\phi_{1}\rangle$ is an eigenvector of $\sigma_{x}$, which means that  the both sides of the inequality (\ref{R1-sigma}) are equal to zero for $|\phi\rangle = |\phi_{1}\rangle$ as it should be in such a case.

 A little more complicated example can be found considering $(3\times3)$ matrices. So, let us consider Gell--Mann matrices $\lambda_{3}, \lambda_{4}$ and $\lambda_{5}$ as an example:
\begin{equation}
\lambda_{3} = \left(
                \begin{array}{ccc}
                  1 & 0 & 0 \\
                  0 & -1 & 0 \\
                  0 & 0 & 0 \\
                \end{array}
              \right),\;\;
\lambda_{4} = \left(
                \begin{array}{ccc}
                  0 & 0 & 1 \\
                  0 & 0 & 0 \\
                  1 & 0 & 0 \\
                \end{array}
              \right),\;\;
\lambda_{5} =   \left(
                \begin{array}{ccc}
                  0 & 0 & i \\
                  0 & 0 & 0 \\
                  -i & 0 & 0 \\
                \end{array}
              \right).
\end{equation}
They are self--adjoint and do not commute,
\begin{equation}
[\lambda_{3},\lambda_{4}]=-i\lambda_{5} \neq  0.
\end{equation}
For these matrices the inequality (\ref{R1}) takes the following form,
\begin{equation}
\Delta_{\psi} \lambda_{3}\,\cdot\,\Delta_{\psi} \lambda_{4} \geq \frac{1}{2}\,\left|\langle [\lambda_{3},\lambda_{4}] \rangle_{\psi} \right| \equiv
\frac{1}{2}\,\left|\langle \lambda_{5} \rangle_{\psi}\right|, \label{R1-lambda}
\end{equation}
where
\begin{equation}
|\psi\rangle = \frac{1}{\sqrt{|a|^{2} + |b|^{2} + |c|^{2}}}\,\left(
                                                         \begin{array}{c}
                                                           a \\
                                                           b \\
                                                           c \\
                                                         \end{array}
                                                       \right), \label{psi-lambda}
\end{equation}
and $a,b,c \in \mathbb{C}$. Now, putting $ a=b=c = 1$ in (\ref{psi-lambda}) one gets
\begin{equation}
|\psi \rangle \;\;\Rightarrow\;\; |\psi_{1}\rangle = \frac{1}{\sqrt{3}}\,\left(
                                                         \begin{array}{c}
                                                           1 \\
                                                           1 \\
                                                           1 \\
                                                         \end{array}
                                                       \right), \label{psi1-lambda}
\end{equation}
which leads to the result $\left|\langle \lambda_{5} \rangle_{\psi_{1}} \right| = 0$, and hence one concludes that for $|\psi_{1}\rangle$ the inequality (\ref{R1-lambda})                                                        takes the following form
\begin{equation}
\Delta_{\psi_{1}} \lambda_{3}\,\cdot\,\Delta_{\psi_{1}} \lambda_{4} \geq 0,\label{R2-lambda}
\end{equation}
exactly as the inequality (\ref{dAdB=0}). More detailed analysis leads to the surprising result: If in (\ref{psi-lambda}) $a=a^{\ast},\,b=b^{\ast},\,c=c^{\ast}$ then there is $\left|\langle \lambda_{5} \rangle_{\psi} \right| = 0$  for any such $a,b,c$. Hence for $|\psi\rangle$ defined by real $a,b,c$ the uncertainty relation (\ref{R1-lambda})  takes the same form as the relation (\ref{dAdB=0}). On the other hand if to consider the more general case when $a,b,c$ are the complex numbers then only for
\begin{equation}
c\, =\, \beta\,a, \label{c=la}
\end{equation}
(where $\beta = \beta^{\ast}\neq0$), one obtains that $\left|\langle \lambda_{5} \rangle_{\psi} \right| = 0$ for any $a$ and $b$ but  $\left|\langle \lambda_{5} \rangle_{\psi} \right| > 0$ for these $a$ and $c$, which do not fulfil the condition (\ref{c=la}) and in this case the uncertainty relation (\ref{R1-lambda}) has the standard form.
Similar examples can be found for self--adjoint matrices or operators acting in any Hilbert space (see, e. g. Sec. 2 in \cite{Dias}).

\section{Particle in the box}

Many similar situations to those discussed in the previous Section, or even paradoxes, can be found when one is analyzing properties of a quantum particle, which  spatial motion is confined to a finite volume. Usually such cases are much more complicated than that discussed in the previous Section. As a  simplest nontrivial example of such a case
the problem of a quantum particle in the box with  perfectly reflecting and impenetrable (rigid) walls will be considered in this Section. We assume that a quantum  non--relativistic particle of mass $m$ is mowing on an interval $(a,b)$ of the real axis. In other words we assume that this particle is in the potential well $V(x)$   defined as follows
\begin{equation}
V(x) = \left\{
         \begin{array}{cl}
           0 & {\rm for}\; \;a < x < b, \\
           +  \infty & {\rm for}\; \;x \leq a \;\;{\rm and }\; \;x \geq b. \\
         \end{array}
       \right. \label{V(x)}
\end{equation}
The hamiltonian $H$, of such a system has a usual form: It is the sum of the kinetic energy, $T$, and the potential $V(x)$: That is $H = T + V(x)$. The
assumed potential $V(x)$ forces the particle to be somewhere between $a$ and $b$. Hence in the position representation the probability  $|\psi (x)|^{2}dx$, (where $\psi(x) = |\psi (x)\rangle$ is the wave function of the particle),  to find this particle having position between $x$ and $x +dx$ out of the interval $(a,b)$  must be zero. Therefore it must be $|\psi (x)|^{2} = 0$ for $x < a$ and $x > b$, and thus within this problem there must be
\begin{equation}
\psi (x) = 0\;\;{\rm for}\;\; x < a \;\;{\rm and}\;\; x >  b. \label{psi=0}
\end{equation}
Taking into account that in this paper we  analyze some properties of the uncertainty relation
our attention will be focussed  only on the operator corresponding to the momentum of the particle considered. In one dimensional models on the real line the position operator $X$ and the momentum operator $P_{x}$ are  self--adjoint operators and when they act in the Hilbert space ${\cal H} = L^{2}(\mathbb{R})$, (where $L^{2}(\mathbb{R})$ denotes the space of square integrable functions on the real line $\mathbb{R}$),  they are defined by $X \psi(x) = x \psi (x)$, (or $X |\psi(x)\rangle  = x |\psi (x)\rangle$),
$P_{x}\phi(x) = - i \hbar \frac{d}{dx} \phi (x)$, (or $P_{x}|\phi(x)\rangle = - i \hbar \frac{d}{dx} |\phi (x)\rangle$) to act on appropriate sets of functions $|\psi (x)\rangle, |\phi (x)\rangle \in L^{2}(\mathbb{R})$. Now if the motion of the particle is confined to a segment $[a,b] \subset \mathbb{R}$, then the suport of the corresponding wave--functions is $[a,b]$ and thus they form a subspace of $L^{2}(\mathbb{R})$, which is identified with the Hilbert space of square integrable functions $L^{2}([a,b])$ on $[a,b]$. The problem is that there is  no a self--adjoint operator acting as $- i \hbar \frac{d}{dx}$ in the subspace of square integrable functions  in $L^{2}([a,b])$ defined by the condition (\ref{psi=0}), that is,  which vanish at the endpoints of the interval $[a,b]$.

Let  us pass now to the analysis of properties of an operator corresponding to the momentum of the particle considered. For simplicity
we will consider the "standard" case when $a=0$ and $b= l> 0$ and the "symmetric" case when $a = -\frac{l}{2}$ and $b = +\frac{l}{2}$ , (see, e.g. \cite{Rob1}).

\subsection{The "standard" case}
Let us consider now
the operator $P_{x}$ in a closed interval $ [0,l]  \ni x $ and let us take for a domain $D(P_{x})$ the following subspace of $L^{2}([0,l])$,
\begin{equation}
D(P_{x}) = \left\{\phi (x), \phi'(x)\in L^{2}([0,l]): \phi(0) = \phi(l)=0 \right\} \label{D(P)}
\end{equation}
where $\phi'(x) = \frac{d}{dx}\phi(x)$. It appears that such defined $P_{x}$ is only a symmetric operator in $D(P_{x})$ but it is not a self--adjoint in  $D(P_{x})$, (see, e.g. \cite{Bon,Garb,Git,Be} and references therein). If one needs a self--adjoint extension of $P_{x}$ then one have to change boundaries defining $D(P_{x})$.
There is a family of self--adjoint extensions of $P_{x}$ "numbered" by a real parameter $\vartheta $, where $0 \leq \vartheta < 2\pi$ \cite{Bon,Garb,Git,Be}, which are denoted as
$P_{x}^{\vartheta}$:
\begin{equation}
 P_{x}^{\vartheta} \phi (x) = - i \hbar \frac{d}{dx} \phi (x), \label{P-s1}
\end{equation}
\begin{equation}
D(P_{x}^{\vartheta}) = \left\{\phi (x), \phi'(x)\in L^{2}([0,l]): \phi(l) = e^{\textstyle{i\vartheta}} \phi(0)\right\}. \label{D(P)theta}
\end{equation}
Note that the set being  the domain $D(P_{x}^{\vartheta}) $ of the operator $P_{x}^{\vartheta}$ is much larger than the set defined in (\ref{D(P)}):
Functions belonging to  $D(P_{x}^{\vartheta})$  do not have to meet the condition $\phi (0) = 0$.
This definition   leads to the following solutions of the eigenvalue problem for $P_{x}^{\vartheta}$:
One finds that the eigenfunctions are
\begin{equation}
u_{n}^{\vartheta}(x) = \frac{1}{\sqrt{l}}\,e^{\textstyle{\frac{i}{\hbar}\,p^{\vartheta}_{n}x}} \label{un-1}
\end{equation}
where $n=0, \pm 1, \pm 2, \ldots$ and the corresponding eigenvalues are:
\begin{equation}
 p^{\vartheta}_{n} = \hbar\,\frac{2 \pi  n + \vartheta}{l}. \label{pn-1}
\end{equation}
For each $\vartheta$ the eigenfunctions $u_{n}^{\vartheta}(x)$ form an orthonormal basis in $L^{2}([0,l])$.
    Let us analyze now the uncertainty relation (\ref{R1}) for the operators $X$ and  $P_{x}^{\vartheta}$. For each $u_{n}^{\vartheta} (x) = | u_{n}^{\vartheta}(x)\rangle$ there is $\Delta_{u_{n}^{\vartheta}}X < l$ and $\Delta_{u_{n}^{\vartheta}}P_{x}^{\vartheta} = 0$. From this one concludes that  there is
    \begin{equation}
    \Delta_{u_{n}^{\vartheta}}X \,\cdot\, \Delta_{u_{n}^{\vartheta}}P_{x}^{\vartheta} = 0,
    \end{equation}
which contradicts (\ref{H1}) and (\ref{[PX]}). This result suggests  that in the case considered there is something wrong with the commutation relation (\ref{[PX]}) and with the modulus of the expectation value of $\langle u_{n}^{\vartheta}(x)|[P_{x}^{\vartheta},X]|u_{n}^{\vartheta}(x)\rangle$. There is
\begin{equation}
\langle u_{n}^{\vartheta}(x)|[P_{x}^{\vartheta},X]|u_{n}^{\vartheta}(x)\rangle = \langle u_{n}^{\vartheta}(x)|P_{x}^{\vartheta}X|u_{n}^{\vartheta}(x)\rangle -
\langle u_{n}^{\vartheta}(x)|XP_{x}^{\vartheta}|u_{n}^{\vartheta}(x)\rangle, \label{u[PX]u}
\end{equation}
and more detailed analysis shows that the position   operator $X$ removes vectors $|\phi (x)\rangle  \in D(P_{x}^{\vartheta})$ from the domain $D(P_{x}^{\vartheta})$ of $P_{x}^{\vartheta}$. Simply, there is $X|\phi (x)\rangle = x |\phi (x)\rangle \stackrel{\rm def}{=} |\chi (x)\rangle$ and, as one can see, the condition
$\chi(l) = e^{\textstyle{i\vartheta}} \chi(0)$ guaranteing that $\chi (x) \in D(P_{x}^{\vartheta})$ can not be fulfilled for such $|\chi (x)\rangle$. This means that the commutator $[P_{x}^{\vartheta}, X]$ does not exist in the considered case (see \cite{Git}). This conclusion  concerns also eigenvectors $u_{n}^{\vartheta}(x)$ of $P_{x}^{\vartheta}$:
The position operator $X$ also removes
vectors $u_{n}^{\vartheta}(x)$ from the domain $D(P_{x}^{\vartheta})$. For every $\chi_{n}^{\vartheta} (x) \stackrel{\rm def}{=} X u_{n}^{\vartheta}(x) \equiv  x u_{n}^{\vartheta}(x)$ one finds that $\chi_{n}^{\vartheta} (l) \equiv l u_{n}^{\vartheta}(l) \neq 0$, whereas
$\chi_{n}^{\vartheta} (0) \equiv 0\cdot  u_{n}^{\vartheta}(0) = 0$ which means that $\chi_{n}^{\vartheta} (x) =  x u_{n}^{\vartheta}(x)$ does not belong to the domain
$D(P_{x}^{\vartheta})$ and therefore the matrix element $\langle u_{n}^{\vartheta}(x)|P_{x}^{\vartheta}X|u_{n}^{\vartheta}(x)\rangle$ is not defined. Hence the relation
(\ref{u[PX]u}) is not defined. This analysis shows that in the considered  "standard" case of the particle, which motion is confined to a segment $[0,l]$,
the uncertainty relation (\ref{H1}) does not hold \cite{Git}.

\subsection{The "symmetric" case}
Let us now analyze  the
symmetric" case
of the particle in the box
when the particle can move only inside the segment $[-\frac{l}{2},\frac{l}{2}]$. In this case
\begin{equation}
V(x) = V^{\ast}(x) = \left\{
         \begin{array}{cl}
           0 & {\rm for}\; \;|x| \leq \frac{l}{2}, \\
           +  \infty & {\rm for}\; \;|x| > \frac{l}{2}. \\
         \end{array}
       \right. \;,\label{V(x)-s}
\end{equation}
the family of self--adjoint extensions $\Pi_{x}^{\vartheta}$ of the operator $P_{x}$ is defined as follows \cite{Al}:
\begin{equation}
 \Pi_{x}^{\vartheta} \phi (x) = - i \hbar \frac{d}{dx} \phi (x),
\end{equation}
\begin{equation}
D(\Pi_{x}^{\vartheta}) = \left\{\phi (x), \phi'(x)\in L^{2}([0,l]): \phi(\frac{l}{2}) = e^{\textstyle{i\vartheta}} \phi(- \,\frac{l}{2})\right\}, \label{D(pi)theta}
\end{equation}
and again $ 0 \leq \vartheta < 2\pi$, $n =0, \pm  1, \pm 2, \ldots$. The solutions of the eigenvalue problem for $\Pi_{x}^{\vartheta}$ have the same form as for the operator
$P_{x}^{\vartheta}$: eigenfunctions are given by (\ref{un-1}) and eigenvalues $p_{n}^{\vartheta}$ are given by the formula (\ref{pn-1}).
Considering the uncertainty relations
(\ref{H1}) and (\ref{R1}) for $X$ and $\Pi_{x}^{\vartheta}$  computed for $|\phi\rangle = |u_{n}^{\vartheta}(x)\rangle$ one finds again that
 $\Delta_{u_{n}^{\vartheta}}X < l$ and $\Delta_{u_{n}^{\vartheta}}\Pi_{x}^{\vartheta} = 0$, which suggest that in the considered case there is
 $\Delta_{u_{n}^{\vartheta}}X \,\cdot\, \Delta_{u_{n}^{\vartheta}}\Pi_{x}^{\vartheta} = 0$ too, which again contradicts (\ref{H1}).
  Now if one wants to verify this conclusion one should use the relation (\ref{R1}), and   then one should to compute the
 expectation value of $\langle u_{n}^{\vartheta}(x)|[\Pi_{x}^{\vartheta},X]|u_{n}^{\vartheta}(x)\rangle$.
 The properties of the matrix element  $\langle u_{n}^{\vartheta}(x)|\Pi_{x}^{\vartheta}X|u_{n}^{\vartheta}(x)\rangle$
 were the crucial
 in the previously considered "standard" case. So, let us analyze the function $\xi_{n}^{\vartheta} (x) \stackrel{\rm def}{=} X|u_{n}^{\vartheta}(x)\rangle$ and let us check if (and when)
 $\xi^{\vartheta}_{n} (x) \in D(\Pi_{x}^{\vartheta})$. There are
 \begin{equation}
 \xi_{n}^{\vartheta}(\frac{l}{2}) = \frac{l}{2}\,u_{n}^{\vartheta}(\frac{l}{2})
 \;\; {\rm and}\;\;
 \xi_{n}^{\vartheta}(- \frac{l}{2}) = -\,\frac{l}{2}\,
 u_{n}^{\vartheta}(- \frac{l}{2}). \label{xi1}
 \end{equation}
 Thus boundaries $\xi_{n}^{\vartheta}(\frac{l}{2} ) = e^{\textstyle{i\vartheta}}\,\xi_{n}^{\vartheta}(- \frac{l}{2} )$ (see (\ref{D(pi)theta})) and properties (\ref{xi1}) leads to the following conclusion:
 In the "symmetric"  case  $\xi^{\alpha}_{n} (x) = x u_{n}^{\alpha}(x) \in D(\Pi_{x}^{\alpha +\pi})$ for all $\alpha$, such that $ 0 < \alpha <\pi$. It is because
 $(-1)$ can be represented by $ e^{\textstyle{i\pi}} \equiv - 1$.  In other words there exists a subfamily of self--adjoint extension of $\Pi_{x}^{\alpha}$, where  $0 < \alpha < \pi$,  such   $\xi^{\alpha}_{n} (x) = x u_{n}^{\alpha}(x) \in D(\Pi_{x}^{\alpha+\pi})$, and in general $X \,D(\Pi_{x}^{\alpha})\, \rightarrow \,D(\Pi_{x}^{\alpha+\pi}) \neq D(\Pi_{x}^{\alpha})$. So, for $0 < \alpha  <\pi$ position operator $X$ moves eigenfunctions of $\Pi_{x}^{\alpha}$ from $ D(\Pi_{x}^{\alpha})$  to domain of $\Pi_{x}^{\alpha +  \pi}$ but nevertheless
$X\,u_{n}^{\alpha}(x) \equiv x u_{n}^{\alpha}(x)\,\not\in D(\Pi_{x}^{\alpha})$ again. For $ \pi \leq \vartheta < 2 \pi$ eigenfunctions of $\Pi_{x}^{\vartheta}$ are removed from any domain of the family of self--adjoint extensions $\Pi_{x}^{\vartheta}$ of the operator $-i\hbar \frac{d}{dx}$.
It is easy to show that $\langle \phi (x)|\Pi_{x}^{\alpha}\,X|\phi(x) \rangle \equiv \langle \phi (x)|\Pi_{x}^{\alpha}\,(X\phi(x) )\rangle \neq
\langle (\Pi_{x}^{\alpha}\phi (x))|(X|\phi(x)) \rangle$ for $|\phi (x) \rangle \in D(\Pi_{x}^{\alpha})$.
This property leads to  a rather unexpected  result that $\langle u_{n}^{\alpha}(x)|[\Pi_{x}^{\alpha},X]|u_{n}^{\alpha}(x)\rangle $ does not exist not only for every $ 0 < \alpha < \pi $ but also for any $\vartheta$, such that $0 < \vartheta < 2\pi$. Note that
if $|\phi (x) \rangle = |u_{n}^{\alpha}\rangle$ then, contrary to the above conclusion,   one expects that $\langle u_{n}^{\alpha}(x)|[\Pi_{x}^{\alpha},X]|u_{n}^{\alpha}(x)\rangle = 0$.
Ignoring the above described subtleties one can see that  in the "symmetric" case the situation is the same as in the "standard" case.
Again the left hand side of the inequality (\ref{R1}) computed for $X$ and $\Pi_{x}^{\vartheta}$ and $|\phi\rangle = |u_{n}^{\vartheta}(x)\rangle$ takes the zero value,
$\Delta_{u_{n}^{\vartheta}}X \,\cdot\,\Delta_{u_{n}^{\vartheta}} \Pi_{x}^{\vartheta} = 0$,
 and the right hand side of (\ref{R1}) does not exists.
 In \cite{Al} a conclusion was that in such a case the momentum is not a physical observable and therefore a consideration of such a case has no a physical justification. As it was said earlier, we analyze properties of uncertainty relations considering them as a mathematical problem and we are interested in finding mathematical solutions of this problem.
 It seems that a solution to a "paradox"
 such "paradoxes can be found
 by carrying out a more detailed analysis of the case under considerations.

From the point of view of the theoretical mechanics
the system considered is the constrained system. Simply here imposed on the positions of the considered particle are restrictions of the geometrical nature, called constraints.
In such a situation the constraint means that certain positions of the particle are permissible and others are forbidden: In the case considered the allowed position, $x$, are: $- \frac{l}{2} \leq x \leq \frac{l}{2}$, whereas
$x <- \frac{l}{2} $ and $x > \frac{l}{2}$ are the forbidden positions. The equation of these constraints can be written as follows:
 \begin{equation}
 |x|\leq \frac{l}{2}
 \end{equation}

It seems that a possible solution to the problem may be
choosing the position operator $X$ in such a way that it would be consistent with constraints.
So, taking into account the constraint equation one can define modified position operator $X_{M}$ acting in $L^{2}(\mathbb{R})$ as follows
\begin{equation}
X_{M} |\psi (x) \rangle = x\, \Theta (\frac{l}{2} + x)\;\Theta (\frac{l}{2} - x)\,|\psi (x) \rangle, \label{X_M}
\end{equation}
\begin{equation}
D(X_{M}) = \left\{\psi (x) \in L^{2}(\mathbb{R}):  \psi (x) = 0\;{\rm for}\; |x | > \frac{l}{2} \right\},\label{DX_M}
\end{equation}
where $\Theta (x)$ is the unit step function: $\Theta (x) = 1 \;{\rm for}\;x \geq 0$ and $\Theta (x) = 0 \;{\rm for}\;x < 0$.
Note that all functions $\phi (x)$ having the set $x \in [-\frac{l}{2}, \frac{l}{2}]$ as a support and belonging to $L^{2}([-\frac{l}{2}, \frac{l}{2}])$ belong also to the domain $D(X_{M})$.
    Using the modified position operator $X_{M}$ one finds for $|\psi (x)\rangle \in L^{2}(\mathbb{R})$ that formally,
    \begin{eqnarray}
    [P_{x}, X_{M}]|\psi (x)\rangle &=&  i \hbar \, \frac{l}{2}\left[ \delta(\frac{l}{2} + x) + \delta (\frac{l}{2} - x)\right] |\psi (x)\rangle \nonumber\\
    && - i \hbar \,\Theta (\frac{l}{2} + x)\;\Theta (\frac{l}{2} - x) |\psi (x)\rangle. \label{[Pi,X]}
    \end{eqnarray}
Note that here operators $P_{x}$ and $X_{M}$ act in $L^{2}(\mathbb{R})$.
  If a segment $[-\frac{l}{2},\frac{l}{2}]$ is the  support of  $|\phi (x) \rangle$ then
 $|\phi (x) \rangle \in L^{2}([-\frac{l}{2}, \frac{l}{2}])$ and  $\Theta (\frac{l}{2} + x)\;\Theta (\frac{l}{2} - x) |\phi (x)\rangle = |\phi (x) \rangle$, which implies that
\begin{eqnarray}
    [P_{x}, X_{M}]|\phi (x)\rangle =  i \hbar \, \frac{l}{2}\left[ \delta(\frac{l}{2} + x) + \delta (\frac{l}{2} - x)\right] |\phi (x)\rangle 
 - i \hbar \, |\phi (x)\rangle. \label{[P,X]a}
    \end{eqnarray}
Hence  for normalized $|\phi (x) \rangle \in L^{2}([-\frac{l}{2}, \frac{l}{2}])$ one obtains that
\begin{equation}
\langle \phi(x)|[P_{x},X_{M}]|\phi (x)\rangle = i \hbar \frac{l}{2} \left[\left||\phi ( - \frac{l}{2})\right|^{2} + |\phi ( \frac{l}{2})|^{2}\right]
- i \hbar.  \label{P,X2}
\end{equation}
The right hand side of (\ref{P,X2}) is zero if
\begin{equation}
 \left|\phi ( - \frac{l}{2})\right| = \left|\phi ( + \frac{l}{2})\right| =  \frac{1}{\sqrt{l}}.  \label{PXm-1}
\end{equation}
 Also the right hand side of (\ref{P,X2}) is zero for every $\phi(x) \in L^{2}([-\frac{l}{2}, \frac{l}{2}])$ such that $\left|\phi(x)\right| = \frac{1}{\sqrt{l}}$.
Note that, among others, all eigenfunctions $u_{n}^{\vartheta}(x)$ (see (\ref{un-1})) of self--adjoint extensions of the momentum operator $P_{x}$ have these properties.
On the other, if
\begin{equation}
\left|\phi ( - \frac{l}{2})\right| = \left|\phi ( + \frac{l}{2})\right| \neq  \frac{1}{\sqrt{l}},\;\;{\rm or,}\;\;\left|\phi ( - \frac{l}{2})\right| \neq \left|\phi ( + \frac{l}{2})\right|, \label{PXm-2}
\end{equation}
then the right hand side of (\ref{P,X2}) is nonzero.

Coming back to the uncertainty relations for the modified position operator $X_{M}$  and momentum $P_{x}$ one finds that
\begin{equation}
\Delta_{\phi} X_{M} \,\cdot\,\Delta_{\phi} P_{x} \geq  \hbar\,\left| \frac{ l}{2} \left[|\phi ( - \frac{l}{2})|^{2} + |\phi ( -\frac{l}{2})|^{2}\right] - 1 \right|.
\label{DX-DP}
\end{equation}
So it seems that the relation (\ref{DX-DP})  is consistent  with the uncertainty relation (\ref{R1}) and the use of the modified position operator may remove inconsistences
with the position--momentum uncertainty relation for particle in the box. Note that, as it was mentioned, the right hand side of the inequality (\ref{DX-DP}) can be zero for vectors satisfying conditions specified after Eq. (\ref{P,X2}).

\section{Uncertainty principles and ${\cal PT}$--symmetric quantum theory}

The uncertainty principle is  one of the most famous predictions of quantum mechanics.
As it was stated on \cite{Mer} (see also \cite{Sen}) {\em "as deduced from within the
quantum theory itself,
the uncertainty principle only
prohibits the possibility of preparing an ensemble of systems
in which all those properties are sharply defined"}. This general statement can be translated for the case of two non--commuting observables $A$ and $B$ as follows:
The possibility of preparing a system, in which the values of observables $A$ and $B$ are sharply defined, can not be realized. This is true within the Schr\"{o}dinger and von Neumann  quantum mechanics. The question is: Is this also true   within the $\cal PT$--symmetric quantum mechanics? Simply, when one goes form the standard (Schr\"{o}dinger and von Neumann) quantum mechanics to $\cal PT$--symmetric quantum mechanics one meets some surprises. One of them is the problem of the uncertainty relations.
In standard quantum mechanics one can ask about
exact values of the position and momentum of the particle independently of that if the Hamiltonian $H$ is known or not and independently of the form of $H$. It is because all observables act in the same, common  Hilbert space $\cal H$ of states and the scalar product in $\cal H$ does not depend on the choice of the Hamiltonian $H$.
The different situation is in $\cal PT$--symmetric quantum mechanics, where the Hamiltonian $H$ and solutions
of the eigenvalue problem for this $H$ determine the space of states and the "scalar product" in this space \cite{Bender}.

Within the $\cal PT$--quantum mechanics the property that non--self--adjoint but $\cal PT$--symmetric Hamiltonians can have the real eigenvalues is used. Here the {\em  $\cal PT$--symmetric Hamiltonian} means that the Hamiltonian $ H$ is requested to satisfy the following condition,
\begin{equation}
H^{PT} \stackrel{\rm def}{=} {\cal PT}\,H\,{\cal PT} \equiv H, \label{h-pt}
\end{equation}
where the operators $\cal P$ and $\cal T$ are defined as follows:
\begin{equation}
{\cal P} x = - x,\;\; {\cal P} p_{x} = - p_{x},\;\;{\cal T}x = x, \;\;{\cal T} p_{x} = - p_{x}, \label{PT1}
\end{equation}
and $x$ and $p_{x}$ denote position and momentum respectively,
\begin{equation}
{\cal P} \phi (x) = \phi(- x),\;\;{\cal T}\phi (x)  = \phi^{\ast}(x). \label{PT2}
\end{equation}
When $\cal T$ acts in the Hilbert space or in a space with sesquilinear form, then $\langle {\cal T}\psi|{\cal T} \phi\rangle = \langle \phi|\psi\rangle$. From these definitions it follows that ${\cal P}^{2} = {\cal T}^{2} = \mathbb{I}$, $[{\cal P}, {\cal T}]=0$ and that $({\cal PT})^{2} = \mathbb{I}$. This means that ${\cal P} = {\cal P}^{-1}$,  ${\cal T} = {\cal T}^{-1}$ and
${\cal PT} = ({\cal PT})^{-1}$. Thus ${\cal P} X  {\cal P} = - X,\;\; {\cal P} P_{x} {\cal P} = - P_{x},\;\;{\cal T} X {\cal T}= X, \;\;{\cal T} P_{x} {\cal T} = - P_{x}$, where $X$ and $P_{x}$ are the standard position and momentum operators.
In analogy to Hermitian quantum mechanics one can  define the
inner product in this case as
\begin{equation}
(\psi,\phi)^{PT} = \int_{-\infty}^{+\infty} [{\cal PT}\psi (x)]\phi(x)\,dx \equiv  \int_{-\infty}^{+\infty}\psi^{\ast} (-x)\phi(x)\,dx, \label{PT-form-1}
\end{equation}
but, unfortunately, then one runs into the problem of having negative norm for some states.
This problem can be solved by introducing a new operator usually called the $\cal C$ operator expressing a symmetry between the positive and negative norm states. Using this $\cal C$ operator we can define the {\em $\cal CPT$ inner product} as follows
\begin{equation}
(\psi,\phi)^{CPT}= \int \psi^{CPT}(x) \phi (x) dx,
\end{equation}
where $ \psi^{CPT}(x) = {\cal C}[{\cal PT} \psi(x)] = \int {\cal C}(x,y) \psi^{\ast} (-y) dy$. This inner product satisfies the requirements for quantum theory defined by $H$ and the norm defined by means of this product is positive.
In order to find within the $\cal PT$--symmetric quantum mechanics a proper space of states with the proper inner product such as   $\cal CPT$ inner product (that is the $\cal C$ operator) one must find solutions of the eigenvalue problem,
\begin{equation}
H \phi_{n}(x) = E_{n} \phi_{n} (x), \label{H-phi}
\end{equation}
 for a given $\cal PT$--symmetric Hamiltonian
 $H^{PT}$.
 If $H^{PT} = H$ then the eigenvalues $E_{n}$ are real.
Having solutions of Eq. (\ref{H-phi}) one can construct a suitable $\cal C$ operator, e.g., as follows \cite{Bender,Bender1,Wang,Moz}
\begin{equation}
{\cal C}(x,y) = \sum_{n=0}^{\infty} \phi_{n}(x)\, \phi_{n}(y).
\end{equation}
Then simply ${\cal C} \phi_{n}(x) = \int {\cal C}(x,y) \phi_{n}(y) dy =  (-1)^{n} \phi_{n}(x)$ (see, e.g. \cite{Bender}).  There are ${\cal P}^{2} = {\cal C}^{2} = \mathbb{I} $, but ${\cal P} \neq {\cal C}$, and $[{\cal P}, {\cal C}] \neq 0$ but $[{\cal C}, {\cal PT}]=0$ and $[{\cal C},H]=0$. The problem is that the calculation of $\cal C$ is very nontrivial for a given $H$: One have to find solutions of the eigenvalue problem for this $H$.
 Having the   $\cal C$ operator one can define observables.

 In ordinary quantum mechanics the condition for a linear operator $A$ to be an observable is that $A$ has to be self--adjoint: $A = A^{+}$. This condition provides the expectation value $\langle \phi|A|\phi\rangle$ of $A$ in a given normalized state $\langle \phi|\phi\rangle =1$, to be real,
 Within the  $\cal PT$--symmetric quantum theory this condition is replaced by the following one: $A^{CPT} \stackrel{\rm def}{=} {\cal CPT}\; A\; {\cal CPT} = A^{T}  $, where $A^{T}$ denotes the transpose of $A$ \cite{Bender}. This  means that if $A$ satisfies  this condition then the expectation value of $A$  calculated for a given state using $\cal CPT$ inner product   is real \cite{Bender} and therefore this operator $A$ can be considered as the observable.
 Note that this condition depends on $\cal C$ and the form of $\cal C$ is determined by solutions of the eigenvalue problem for $H$. Hence the inner product $(.,.)^{CPT}$ depends on the choice of $H$. So, in general it may happen that an linear operator $A$  satisfies the condition ${\cal CPT}\; A \; {\cal CPT} = A^{T} = $ for $H$ but it does not satisfy analogous condition for problem described by a Hamiltonian  $H_{1} \neq H$
 (assuming that $H$ and $H_{1}$ does not have common eigenfunctions).
 Every $\cal CP$--symmetric Hamiltonian $H$ satisfies the condition ${\cal CPT}\, H\,{\cal CPT} = H^{T} = H$, so the Hamiltonian $H$ is an observable. Now having observables and expectations values one can think about uncertainty relations. It turns out that in typical models considered in $\cal PT$--symmetric quantum mechanics the position $x$ and momentum $p$ are not observables (sse, e.g. \cite{Bender}). Simply in these models, e.g. the expectation value of $x$ in the ground state is a negative imaginary number as it was shown in \cite{Bender}. Thus there is no position  operator in $\cal PT$--symmetric quantum mechanics \cite{Bender}.
  This means that there is no a place  for the Heisenberg uncertainty relation (\ref{H1}) in $\cal PT$--symmetric quantum mechanics.
 So, the question arises: can the system be prepared in a state, in which the position and momentum are sharply defined in such cases?
    Of course, within the $\cal PT$-- symmetric quantum mechanics one can try to find  two non commuting observables $A$ and $B$, that is, such linear operators that
${\cal CPT}\; A\;  {\cal CPT} = A^{T}  $ and ${\cal CPT}\; B\; {\cal CPT} = B^{T} $, and to derive a relation corresponding to the uncertainty relation (\ref{R1}), but such a relation can never be considered to be universally valid. It is because the operator $\cal C$, the inner product,  $(.,.)^{CPT}$, in the state space and thus geometry of this state space are determined  by a given Hamiltonian $H$ for the problem considered.
In conclusion
one may wonder if it makes sense to ask about uncertainty relations in $\cal PT$--symmetric quantum mechanics.

In the light of the consequences of the $\cal PT$--symmetric quantum mechanics and of the fact that within the standard quantum mechanics uncertainty relations only results from  the geometry of the state space, the question concerning uncertainty relations may arise:  Are they the intrinsic and inherent property of the quantum systems, or maybe,  are they a result of our choice of the state space? Taking into account applications of the $\cal PT$--symmetric Hamiltonians in quantum field theory, quantum optics, in condensed matter physics, etc., and the reported result in \cite{Roz,Bru},
where a violation of Heisenberg’s
{\em "measurement--disturbance relationship"} was observed,  this question seems to be nontrivial and important.

\section{Discussion and conclusions}

As it was mentioned in the Introduction, There is almost common
belief that having a pair of non--commuting self--adjoint operators (observables) $A$ and $B$ one always finds that the product of standard deviations $\Delta_{\phi}A$ and
$\Delta_{\phi}B$ calculated for them, (where $|\psi\rangle \in {\cal H}$), is always larger than some nonzero positive number:
$\Delta_{\phi} A \,\cdot\, \Delta_{\phi} B \geq a >0$.
In Sections 3 --- 5 it was shown that  such a belief may lead to confusions. As it was shown in Section 3, in 2--dimensional, or 3--dimensional state spaces there are many examples of self--adjoint matrices (operators) and vectors in state spaces  such that the product of the standard deviations calculated for them is greater than or equal to zero. Similar cases can be found in n--dimensional state spaces. These observations seem to be highly non--trivial in the case of studying the properties of two--, three--, and n--level quantum systems, which have many applications and which are  intensively studied in the context of applications, e.g. in the theory of quantum computers, and in another cases.
Simply, the examples presented in Sec. 3 show that the uncertainty principle (\ref{R1}) may not work in many cases in n--level systems, although at first glance it seems it must work. This means that in order to avoid unpleasant surprises, when examining such systems and drawing general conclusions from them based on the uncertainty principle, one must carefully check each such case.

Similar observations concern also, e.g  systems having the space $L^{2}(\mathbb{R})$, or $L^{2}([a,b]) \subset L^{2}(\mathbb{R})$ as a state space. Examples of such a situations has been studied in Sec. 4. In this Section the attention was focused on the standard Heisenberg position--momentum uncertainty relation (\ref{H1}) for a quantum particle in the box with  perfectly reflecting and impenetrable walls. The detailed description of this problem can be found, e.g. in \cite{Rob1,Bon,Git,Be,Al} and this is why we do not analyze all the details and subtleties of this problem, but focus our attention on the momentum of the particle considered. Much more details concerning this momentum can be found, e. g. in \cite{Git} and also in the nice paper \cite{Garb}. In Subsection 4.1 the "standard" case of a particle in the box has been considered, when the potential $V(x)$ is given by formula (\ref{V(x)}) with $a=0$ and $b=l$ and the state space is $L^{2}([0,l])$. Analyzing the position--momentum uncertainty relation in this case the family of self--adjoint adjoint extensions of the momentum operator $P_{x}^{\vartheta}$ (see (\ref{P-s1}), (\ref{D(P)theta})) was used to find the uncertainty relation. Unfortunately, a naive direct use of the relation (\ref{R1}) to find the suitable relation leads to paradoxical situation, where the left--hand side of the relation (\ref{R1}) is zero for eigenvectors of $P_{x}^{\vartheta}$, $|\phi\rangle = |u_{n}^{\vartheta}(x)\rangle$, whereas, according to  (\ref{[PX]}) the right hand side is non--zero. A more detailed analysis  shows that the position operator $X$ removes for any $\vartheta$ vectors $|u_{n}^{\vartheta}(x)\rangle$ from the domain, $D(P_{x}^{\vartheta})$,
of the operator $P_{x}^{\vartheta}$, which means that  the commutator $[X,P_{x}^{\vartheta}]$ does not exist in this case and therefore the right hand side of the inequality (\ref{R1}) does not exits. What is more, it appears that for any $\vartheta$ the position operator $X$ removes also from $D(P_{x}^{\vartheta})$ all vectors such that $\phi (0) \neq 0$, which has a consequence that in the case of these vectors the commutator $[X,P_{x}^{\vartheta}]$ can not be calculated.
As a result the position--momentum uncertainty relation can not be derived from (\ref{R1}) in the mentioned cases (see also \cite{Git}).

 A slightly different picture one meets in the "symmetric" case of the particle in the box discussed in Subsection 4.2. Here the Hilbert space $L^{2}([-\frac{l}{2}, \frac{l}{2}]) \subset L^{2}(\mathbb{R})$ is the state space
 and for $ 0 \leq \vartheta \equiv \alpha +  \pi < 2\pi$ the position operator $X$ moves vectors $|\phi (x) \rangle \in D(\Pi_{x}^{\alpha})$  from the domain of the self--adjoint extension $\Pi_{x}^{\alpha}$ of the momentum operator to the domain $D(\Pi_{x}^{\alpha + \pi}) \neq D(\Pi_{x}^{\alpha})$ of the operator $\Pi_{x}^{\alpha + \pi}$. Unfortunately this means that  for $|\phi(x) \in  D(\Pi_{x}^{\alpha})$ vectors $X |\phi (x)\rangle \in  D(\Pi_{x}^{\alpha + \pi}) \neq  D(\Pi_{x}^{\alpha})$ and therefore
 the matrix element of the commutator $\langle \phi(x)|[\Pi_{x}^{\alpha},\,X ]|\phi (x)\rangle $ calculated for $|\phi (x) \rangle = |u_{n}^{\alpha}(x)\rangle$, (where $|u_{n}^{\alpha}(x)\rangle$ is an eigenvector of $\Pi_{x}^{\alpha}$) can not exist. So, as the result one can not calculate the unceratinty relation (\ref{R1}) for the position and momentum.

It seems that the root cause of these inconsistencies and paradoxes is the quantization procedure for particles whose spatial motion is confined to a finite volume.
The standard canonical quantization procedure leads to the correct results if conventional space--phase variables $p_{x}$, (momentum), and $x$, (position), can vary from $-\infty$ to $+\infty$: $ |p_{x}| < \infty,\; |x| < \infty$. As it is stated in \cite{Klauder}: {\em Conventional phase--space variables, such as $p$ and $q$, where $-\infty < p, q <+\infty$,
with Poisson brackets $\{q, p\} = 1$, are natural candidates to promote to basic quantum operators in the procedures that canonical quantization employs}. Simply if the spatial motion of the particle is confined to a finite volume then we have the constrained system and the quantization procedure should take into account this fact and to be consistent with the constraints.

As it has been shown the paradox appearing in the "symmetric" case can be removed
if to use the observation that from the classical point of view the particle in the box is the system with constraints and to use the modified position operator, $X_{M}$, (defined by (\ref{X_M}), (\ref{DX_M})) consistent with these constraints and replacing the standard position operator $X$. The use of the operator $X_{M}$ changes  the commutator (\ref{[PX]}) giving the results (\ref{[Pi,X]}), (\ref{[P,X]a}) and (\ref{P,X2}).
Applying the modified position operator to calculate $\Delta_{\phi}X_{M}$ for $\phi (x) \in L^{2}(-[\frac{l}{2},\frac{l}{2}])$ and inserting  the commutators  (\ref{[P,X]a}) into the right hand side of the inequality (\ref{R1}) may remove the above described inconsistencies  appearing in the "symmetric" case considered in Subsection 4.2.
Simply using the modified position operator $X_{M}$ and the commutator  (\ref{[Pi,X]}), (\ref{[P,X]a}), (\ref{P,X2}) one finds that expected value $\langle[P_{x},X]\rangle_{\phi}$ of the commutator $[P_{x},X]$ equals zero for $\phi (x) = u_{n}^{\alpha}(x)$ and also for $\phi (x)$ fulfilling the condition (\ref{PXm-1}) and that described below this formula. This commutator is nonzero for $\phi (x)$ satisfying conditions (\ref{PXm-2}).
Summarizing this part, it should be noted that the properties of the modified position operator $ X_{M}$  defined in Sec. 4.2  and its implications are a proposal that requires further in-depth studies.

One more observation concerning the "symmetric" case.
It appears the in this case
the potential $ V (x)$ is not only symmetric with respect to the origin of the coordinate system but also with respect to the combined transformations of the space reflection, $\cal P$ and the inversion of time, $\cal T$, which are  defined by Eqs (\ref{PT1}), (\ref{PT2}).
The potential $V(x) $, and also eigenfunctions, $ u_{n}^{\vartheta}(x)$ of the operator $\Pi_{x}^{\vartheta}$ and the domain, $D(\Pi_{x}^{\vartheta})$, are invariant under the $\cal PT$ transformation, which may explain
slight differences between "standard" and "symmetric" cases of the particle in the box.

In general,  the problem of the particle in the infinite square well has not only a long tradition of illustrating quantum concepts but also it has important practical meaning.
Full and accurate knowledge of the properties of the particle in the potential well is necessary to understand the properties of such systems as quantum dots, quantum traps and and related problems. A problem of a single slit diffraction experiment and  the uncertainty relation of position and momentum in such a system, where the spatial dimension is one
($x \in [-\frac{\Delta x}{2}, \frac{\Delta x}{2}]$ --- see \cite{Schu}) is an example of such related problems: In \cite{Schu} the uncertainty relation was evaluated  for a subset of functions with support in $[-\frac{\Delta x}{2}, \frac{\Delta x}{2}]$, which satisfy Dirichlet conditions at $x = -\frac{\Delta x}{2}$ and $x = + \frac{\Delta x}{2}]$.
An another related problem is a problem of the uncertainty principle for  a particle localized in a compact domain $D \subset \mathbb{R}^{n}$ considered in  \cite{Schu1}, where the approach used in  \cite{Schu} was applied.
In this context, the information on the behavior of  a particle resulting from the uncertainty principles seems to be of key importance for a full understanding of all the effects occurring in such systems and as it has been shown in Sec. 4 it is very nontrivial problem (see also, e.g. \cite{Al}) and still needs  further studies.

The detailed and rigorous mathematical analysis of the Heisenberg's relation (\ref{H1}) together with (\ref{R1}) shows that, e. g. for observables $A \stackrel{\rm def}{=} X^{n}$ and $B \stackrel{\rm def}{=} P^{m}$, (where $P = - i\hbar \frac{d}{dx}$ and $m,n \in \mathbb{N}$),
using the so--called unitary dilation operator
one can build
from a normalized state $|\psi (x)\rangle \in L^{2}(\mathbb{R})$
such a function
that the product of standard deviations of $X^{n}$ and $P^{m}$ calculated for this function  can vanish (for details see, e. g. \cite{Dias}). This suggest that relations (\ref{H1}), (\ref{R1}) may   not be good relations, strictly speaking that the product $\Delta_{\psi}A\,\cdot\,\Delta_{\psi}B$ may  not be a good measure of the uncertainty. This is why in many papers were considered other relations between standard deviations $\Delta_{\psi}A$ and $\Delta_{\psi}B$ \cite{Fol,Cow} having, e. g. a form of a sum of the squares of the standard deviations (see e. g. \cite{Fol}): $\| \delta_{0} A |\psi \rangle\|^{2} + \| \delta_{0} B |\psi \rangle\|^{2} \geq c_{0} > 0$, where $c_{0}$ is a real number and $\delta_{0}A,\;\delta_{0}B$
denote a suitably rescaled deviations $\delta A$ and $\delta B$ to have the same dimension, or to be dimensionless. A simple example of such a relation can be found analyzing the case of Pauli matrices considered  in Sec. 3: It is enough to take a sum of  squares of $\Delta_{\phi}\sigma_{x}$ and $\Delta_{\phi}\sigma_{y}$.
As it is seen, the inconsistencies of this type and others discussed in previous Sections are integrated into inequality (\ref{R1}).
For this reason, attempts are being made to improve and refine the Heisenberg  as well as Robertson and Schrodinger uncertainty relations (see, e. g. \cite{Dias,Fol,Cow,Dou,Park}).

From the analysis presented in Sec. 5 it follows that a status and role of the uncertainty relations
(\ref{ur-c}), (\ref{H1}), (\ref{R1})
in $\cal PT$--symmetric quantum theory seems to be unclear. It is because the definition of the observable is determined by the choice of $\cal PT$--symmetric Hamiltonian $H$. This means that, for example, if $A$ and $  B$ are observables
with respect to the inner product $(.,.)^{C_{1}PT}$ defined by means of the eigenfunction of the $\cal PT$--symmetric Hamiltonian $H_{1}$, from which the operator ${\cal C}_{1}$ is build,  then they need not  be observables with respect to the inner product $(.,.)^{C_{2}PT}$ defined by eigenvectors of
such $H_{2} \neq H_{1}$
that $H_{1}$ and $H_{2}$ have not common eigenfunctions.
Hence the relation corresponding to the uncertainty relation (\ref{R1}) can not be considered as universally valid: The relations derived for $H_{1}$ need not hold within $\cal PT$--symmetric quantum mechanics generated by the $\cal PT$--symmetric Hamiltonian $H_{2} \neq H_{1}$. What is more, as it was stated in \cite{Bender}, in typical models consider4d within $\cal PT$--symmetric
quantum mechanics the position and momentum are not observables. This means that the standard position--momentum uncertainty relation (\ref{H1}) can not be derived in such cases. In conclusion: Within the $\cal PT$--symmetric quantum mechanics the problem of relations corresponding to the uncertainty relation (\ref{R1}) is open and needs further studies.\\
\hfill\\

\noindent
{\bf Acknowledgments}

This work was supported by
the program of the Polish Ministry
of Science and Higher Education under the name "Regional
Initiative of Excellence" in 2019 --- 2022, Project No. 003/RID/2018/19; (Funding amount: 11 936 596.10 PLN).\\
\hfill\\
{\bf The author contribution statement:} The author declares that there are no conflicts of interest
regarding the publication of this article  and that all results presented in this article are the author's own results.

\end{document}